\documentclass[aps,twocolumn,showpacs,footinbib]{revtex4-1}%
\usepackage{amssymb}
\usepackage{amsmath}
\usepackage{amscd}
\usepackage{graphicx}
\usepackage{subfigure}
\usepackage{float}
\begin{document}

\title[Short Title]{Steady-State Entanglement for Distant Atoms by Dissipation in Coupled Cavities}

\author{Li-Tuo Shen$^{1}$}
\author{Xin-Yu Chen$^{1}$}
\author{Zhen-Biao Yang$^{2}$}
\author{Huai-Zhi Wu$^{1}$}
\author{Shi-Biao Zheng$^{1}$}
\email{sbzheng@pub5.fz.fj.cn}

\affiliation{$^{1}$Lab of Quantum Optics, Department of Physics,
Fuzhou University, Fuzhou 350002, China\\$^{2}$Key Laboratory of
Quantum Information, University of Science and Technology of China,
CAS, Hefei 230026, China}

\begin{abstract}
We propose a scheme for the generation of entangled states for two
atoms trapped in separate cavities coupled to each other. The scheme
is based on the competition between the unitary dynamics induced by
the classical fields and the collective decays induced by the
dissipation of two delocalized field modes. Under certain
conditions, the symmetric or asymmetric entangled state is produced
in the steady state. The analytical result shows that the
distributed steady entanglement can be achieved with high fidelity
independent of the initial state, and is robust against parameter
fluctuations. We also find out that the linear scaling of
entanglement fidelity has a quadratic improvement compared to
distributed entangled state preparation protocols based on unitary
dynamics.
\end{abstract}

\maketitle


There have been various practical applications for quantum entangled
states, ranging from quantum teleportation
\cite{PR1935-47-777,JMO1993-40-1195} to universal quantum
computation \cite{PRL2000-85-2392,cup2000}. The main obstacle in
preserving entanglement is decoherence induced by the environment.
Recently, dissipative state preparation has become a focus in
quantum computation and entanglement engineering
\cite{PRL2011-106-090502,arXiv1110.1024v1,PRA2011-84-022316,arXiv1005.2114v2,
PRA2011-83-042329,PRL2011-107-120502,PRA2010-82-054103,
PRA2007-76-062311,
EPL-85-20007,PRL-89-277901,PRL-91-070402,PRA-76-022312,PRE-77-011112,JPA-39-2689,
PRA-77-042305,PRL-100-220401}, which uses decoherence as a
powerful resource without destroying the quantum entanglement. These
schemes are robust against parameter fluctuations, obtain high
fidelity entanglement with arbitrarily initial states, and do not
need accurate control of the evolution time. Particularly,
Kastoryano and Reiter \emph{et al.} \cite{PRL2011-106-090502,arXiv1110.1024v1} proposed a novel
scheme for dissipative preparation of entanglement for two atoms in
an optical cavity which gets a qualitative improvement in the
scaling of the fidelity with optimal cavity parameters as compared
to any state preparation protocol with coherent unitary dynamics.
However, most of the previous theoretical schemes and experiments
\cite{PRL2011-107-080503} concentrate on the case in which two atoms
are trapped in a single cavity.

For distributed quantum information processing, it is a basic
requirement to perform state transfer and quantum gate operation
between separate nodes of a quantum network. To overcome the
difficulty of individual addressability existing in a single cavity,
efforts have been devoted to the coupled-cavity models both
theoretically \cite{PRA2008-78-063805,LPR2008-2-527,PRA-79-050303,
PRA-78-022323,PRA-76-031805R,Nature2006-2-849,Nature2006-2-856} and
experimentally \cite{Nature2003-421-925}. Most works
on the coupled-cavity system focused on the traditional coherent
unitary dynamics, requiring precise timing and special initial
states. Clark \emph{et al.} \cite{PRL2003-91-177901} proposed a
scheme to entangle the internal states of atoms in separate optical
cavities using technique of quantum reservoir engineering, however
the scheme requires a complex atomic level configuration.
Furthermore, the evolution towards the steady state slows down as the entanglement
of the desired state increases.

In this paper, we generalize the idea of Refs.
\cite{PRL2011-106-090502,arXiv1110.1024v1} and propose a scheme for producing
distributed entanglement for two atoms trapped in coupled cavities.
Due to the coherent photon hopping between the two cavities, the system
is mathematically equivalent to that involving two atoms collectively interacting
with two common nondegenerate field modes symmetrically and asymmetrically,
respectively. Each delocalized field mode induces a collective atomic decay channel.
The present scheme uses the competition between the transitions induced by the
microwave fields and the two collective atomic decay channels to
drive atoms to a symmetric or asymmetric entangled state. Analytical
and numerical results show that the distributed steady entanglement
can be obtained with high fidelity. The scheme is independent of the
initial state and robust against parameter fluctuations. No photon
detection, or unitary feedback control is required. The linear
scaling of $F$ is a quadratic improvement on the cooperativity
parameter $C^{-1}$ compared to any known entangled state preparation
protocol for coupled-cavity systems
\cite{PRL2003-91-177901,Nature2003-421-925,Nature2006-2-849,Nature2006-2-856,OC2010-283-3052}, whose
optimal value is $1-F$ $\varpropto$ $C^{-1/2}$.

The experimental setup, as shown in Fig. 1, consists of two
identical $\Lambda$-type atoms each having two ground states
$|0\rangle$ and $|1\rangle$, and an excited state $|2\rangle$ and
trapped in one detuned cavity. An off-resonance optical laser with
detuning $\Delta$ drives the transition $|0\rangle$
$\leftrightarrow$ $|2\rangle$ and a microwave field resonantly
drives the transition $|0\rangle$ $\leftrightarrow$ $|1\rangle$. The
cavity mode is coupled to the $|1\rangle$ $\leftrightarrow$
$|2\rangle$ transition with the detuning $\Delta-\delta$, where
$\delta$ is the cavity detuning from two photon resonance. We here
assume a $\theta_{M}$ phase difference between the microwave fields
applied to the two atoms. Under the rotating-wave approximation, the Hamiltonian of the whole
system in the interaction picture reads $H_{I}$ = $H_{0}$ + $H_{g}$
+ $V_{+}$ + $V_{-}$, where
\begin{eqnarray}\label{e1-e3}
H_{0}&=&\delta
(a_{1}^{\dag}a_{1}+a_{2}^{\dag}a_{2})+\Delta(|2\rangle_{1}\langle2|+|2\rangle_{2}\langle2|)\cr&&
+[g|2\rangle_{1}\langle1|a_{1}+g|2\rangle_{2}\langle1|a_{2}+H.c.]\cr&&
+J(a_{1}^{\dag}a_{2}+a_{1}a_{2}^{\dag}),\\
H_{g}&=&\frac{\Omega_{M}}{2}(e^{i\theta_{M}}|1\rangle_{1}\langle0|+|1\rangle_{2}\langle0|)+H.c.,\\
V_{+}&=&\frac{\Omega}{2}(|2\rangle_{1}\langle0|+|2\rangle_{2}\langle0|),
\end{eqnarray}
$V_{-}=(V_{+})^{\dagger}$, $a_{i}$ is the cavity field operator in
cavity $i$ ($i=1,2$), $J$ is the photon-hopping strength which
describes cavity and cavity coupling, $g$ is the atom-cavity
coupling constant, $\Omega$ and $\Omega_{M}$ represent the classical
laser driving strength and the microwave driving strength,
respectively. $\theta_{M}$ = $\pi$ (or $0$) guarantees a high
fidelity for asymmetric steady-state $|S\rangle$ $=$
$(|01\rangle-|10\rangle)/\sqrt{2}$ $($ or symmetric steady-state
$|T\rangle$ $=$ $(|01\rangle+|10\rangle)/\sqrt{2}$ $)$. Let us
introduce two delocalized bosonic modes $c_{1}$ and $c_{2}$, and
define asymmetric mode $c_1$ = $(a_1-a_2)/\sqrt{2}$ and symmetric
mode $c_2$ = $(a_1+a_2)/\sqrt{2}$, which are linearly related to the
field modes of two cavities. In terms of the new operators, the
Hamiltonian $H_{0}$ can be rewritten as
\begin{eqnarray}\label{e4}
H_{0}&=& \frac{g}{\sqrt{2}}[|2\rangle_{1}\langle1|(c_1+c_2)
+|2\rangle_{2}\langle1|(c_2-c_1)+H.c.]\cr&&+(\delta-J)c_{1}^{+}c_{1}+(\delta+J)c_{2}^{+}c_{2}
+\Delta\sum_{i=1,2}|2\rangle_{i}\langle2|.
\end{eqnarray}
\begin{figure}
\includegraphics[width=1\columnwidth]{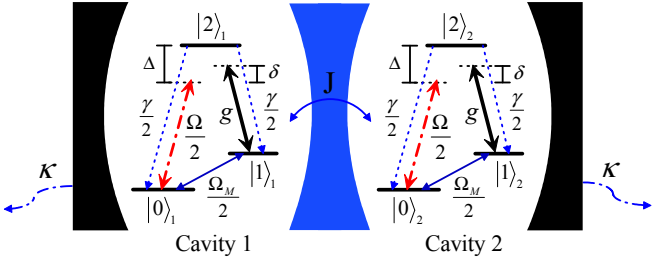} \caption{(Color
online) Experimental setup for dissipative preparation of entangled
steady-state between two $\Lambda$-type atoms trapped in two coupled
cavities. The atom in each detuned cavity has two ground states
$|1\rangle$ and $|0\rangle$, and one excited state $|2\rangle$,
which is driven by the same off-resonance optical laser. The
microwave fields applied to the two atoms differ by a relative phase
of $\theta_{M}$.}
\end{figure}

The Hamiltonian $H_{0}$ describes the asymmetric coupling for the two
atoms to the delocalized field mode $c_1$ and the symmetric coupling
to $c_2$. Due to the photon hopping these two delocalized field modes
are nondegenerate and each induces a collective atomic decay channel.
The photon decay rate of cavity $i$ ($i$ = $1,2$) is denoted as
$\kappa_{i}$ and the spontaneous emission rate of the atoms is
denoted as $\gamma_j$ ($j$ = $1,2,3,4$). Under the condition
$\kappa_1$ = $\kappa_2$ =$\kappa$, the Lindblad operators associated
with the cavity decay and atomic spontaneous emission can be
expressed as $L^{\kappa_1}$ = $\sqrt{\kappa}$ $c_1$, $L^{\kappa_2}$
= $\sqrt{\kappa}$ $c_2$, $L^{\gamma_1}$ = $\sqrt{\gamma_1}$
$|0\rangle_{1}\langle2|$, $L^{\gamma_2}$ = $\sqrt{\gamma_2}$
$|0\rangle_{2}\langle2|$, $L^{\gamma_3}$ = $\sqrt{\gamma_3}$
$|1\rangle_{1}\langle2|$, $L^{\gamma_4}$ = $\sqrt{\gamma_4}$
$|1\rangle_{2}\langle2|$. We assume $\gamma_1$ = $\gamma_2$ =
$\gamma_3$ = $\gamma_4$ = $\gamma/2$ for simplicity.

Under the condition of weak classical laser field, we can
adiabatically eliminate the excited cavity field modes and excited
states of the atoms when the excited states are not initially
populated. To tailor the effective decay processes to achieve a
desired steady-state, we introduce an effective operator formalism
based on second-order perturbation theory
\cite{PRL2011-106-090502,arXiv1110.1024v1,arXiv:1112.2806v1}. Then
the dynamics of our coupled cavity system is governed by the
effective Hamiltonian $H_{eff}$ and effective Lindblad operator
$L_{eff}^{x}$
\begin{eqnarray}\label{e5-e6}
H_{eff}&=&-\frac{1}{2}V_{-}[H^{-1}_{NH}+(H^{-1}_{NH})^{\dag}]V_{+}+H_{g},\\
L_{eff}^{x}&=&L^{x}H^{-1}_{NH}V_{+},
\end{eqnarray}
where $H^{-1}_{NH}$ is the inverse of the non-Hermitian Hamiltonian
$H_{NH}$ = $H_{0}$ $-$ $\frac{i}{2}\sum_{x}(L^{x})^{\dag}L^{x}$.
\begin{figure}
\centering
\includegraphics[width=1\columnwidth]{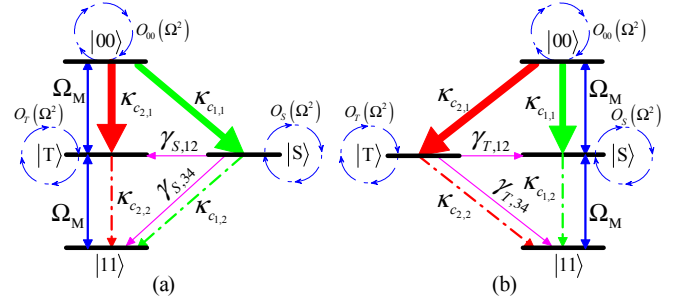} \caption{(Color
online) Two effective models for coherent and dissipative
interactions among states $|00\rangle$, $|S\rangle$ ($|T\rangle$)
and $|11\rangle$, where two microwave fields cause rapid
transitions. The atoms decay through the cavity from $|00\rangle$ to
$|T\rangle$ and $|S\rangle$ with effective decay rates
$\kappa_{c_{1,1}}$ and $\kappa_{c_{2,1}}$, and from $|T\rangle$ and
$|S\rangle$ to $|11\rangle$ with the effective decay rates
$\kappa_{c_{1,2}}$ and $\kappa_{c_{2,2}}$, where $\kappa_{c_{1,1}}$
$\gg$ $\kappa_{c_{1,2}}$ and $\kappa_{c_{2,1}}$ $\gg$
$\kappa_{c_{2,2}}$. $\gamma_{S,12}$, $\gamma_{S,34}$,
$\gamma_{T,12}$ and $\gamma_{T,34}$ are the effective spontaneous
emission rates. The loop-like element $O_{X}(\Omega^2)$ ($X=00,S,T$)
represents the square of the coefficient in the corresponding term
$|X\rangle\langle X|$ within $H_{eff}$ without considering $H_{g}$.
(a) $\theta_{M}=0$. (b) $\theta_{M}=\pi$.}
\end{figure}
The resulting effective master equation in Lindblad form is
\begin{eqnarray}\label{e7-e10}
\dot{\rho}&=&i[\rho,H_{eff}]+\sum_{x}\{L_{eff}^{x}\rho
(L_{eff}^{x})^{\dag}-\frac{1}{2}[(L_{eff}^{x})^{\dag}L_{eff}^{x}\rho\cr&&+\rho(L_{eff}^{x})^{\dag}L_{eff}^{x}]\},\\
H_{eff}&=&-Re[\frac{\Omega^2}{4}\widetilde{R}_{3}]|S\rangle\langle
S|-Re[\frac{\Omega^2}{4}\widetilde{R}_{2}]|T\rangle\langle
T|\cr&&-Re[\frac{\Omega^2}{2}\widetilde{R}_{1}]|00\rangle\langle00|+H_{g},\\
L_{eff}^{\kappa_{1}}&=&\sqrt{\frac{(\delta+J)^2g_{eff}^2\kappa/4
}{A_{\kappa_{1}}^2+B_{\kappa_{1}}^2}}|S\rangle\langle00|
+\sqrt{\frac{g_{eff}^2\kappa/4}{C_{\kappa_{1}}^2+D_{\kappa_{1}}^2}}|11\rangle\langle
S|,\cr&&\\
L_{eff}^{\kappa_{2}}&=&\sqrt{\frac{(\delta-J)^2g_{eff}^2\kappa/4
}{A_{\kappa_{2}}^2+B_{\kappa_{2}}^2}}|T\rangle\langle00|
+\sqrt{\frac{g_{eff}^2\kappa/4}{C_{\kappa_{2}}^2+D_{\kappa_{2}}^2}}|11\rangle\langle
T|,\cr&&
\end{eqnarray}
where $Re[$ $]$ denotes the real part of the argument,
\begin{eqnarray}\label{e11}
g_{eff}&=&\frac{g\Omega}{\Delta},
\delta^{'}=\delta-\frac{i}{2}\kappa,\Delta^{'}=\Delta-\frac{i}{2}\gamma,\cr\cr
\widetilde{R}_{1}&=&\frac{-(\delta^{'}-J)(\delta^{'}+J)}{\delta^{'}g^2-\Delta^{'}(\delta^{'}-J)(\delta^{'}+J)},\cr
\widetilde{R}_{2}&=&\frac{-gJ-\delta^{'}g^2+\Delta^{'}(\delta^{'}-J)(\delta^{'}+J)}{[g^2-\Delta^{'}(\delta^{'}-J)][g^2-\Delta^{'}(\delta^{'}+J)]},\cr
\widetilde{R}_{3}&=&\frac{gJ-\delta^{'}g^2+\Delta^{'}(\delta^{'}-J)(\delta^{'}+J)}{[g^2-\Delta^{'}(\delta^{'}-J)][g^2-\Delta^{'}(\delta^{'}+J)]},\cr
 A_{\kappa_{1}}&=&A_{\kappa_{2}}=\frac{\delta g^2}{\Delta}-(\delta^2-J^2),\cr
  B_{\kappa_{1}}&=&B_{\kappa_{2}}=\kappa(\delta-\frac{g^2}{2\Delta})+\frac{\gamma(\delta^2-J^2)}{2\Delta},\cr
C_{\kappa_{1}}&=&\frac{g^2}{\Delta}-(\delta-J),D_{\kappa_{1}}=\frac{\kappa}{2}+\frac{\gamma(\delta-J)}{2\Delta},\cr
C_{\kappa_{2}}&=&\frac{g^2}{\Delta}-(\delta+J),D_{\kappa_{2}}=\frac{\kappa}{2}+\frac{\gamma(\delta+J)}{2\Delta}.
\end{eqnarray}

As shown in Fig. 2 (a) and (b), the loop-like elements
$O_{00}(\Omega^2)$, $O_{T}(\Omega^2)$ and $O_{S}(\Omega^2)$
represent the effective-Hamiltonian evolution in three triplet
states $|00\rangle$, $|T\rangle$ and $|S\rangle$ without microwave
fields, respectively. For weak optical driving $\Omega$, $H_{eff}$ $\simeq$ $H_g$.
There exist two effective decay channels characterized by $L_{eff}^{\kappa_{1}}$ and
$L_{eff}^{\kappa_{2}}$ through the two delocalized bosonic modes
$c_{1}$ and $c_{2}$ as compared with the case of Ref.
\cite{PRL2011-106-090502} in which only one decay channel is
mediated. It is the photon hopping that lifts the degeneracy of
the two delocalized field modes and leads to the two independent decay
channels. $L_{eff}^{\kappa_{1}}$ indicates the effective decay
from $|00\rangle$ to $|S\rangle$ at a
rate $\kappa_{c_{1,1}}$ and from $|S\rangle$ to $|11\rangle$ at a
rate $\kappa_{c_{1,2}}$ caused by asymmetric $c_1$ mode, and
$L_{eff}^{\kappa_{2}}$ denotes the effective decay from $|00\rangle$
to $|T\rangle$ at a rate $\kappa_{c_{2,1}}$ and from $|T\rangle$ to
$|11\rangle$ at a rate $\kappa_{c_{2,2}}$ caused by symmetric $c_2$
mode simultaneously. The decay rates $\kappa_{c_{1,1}}$
($\kappa_{c_{1,2}}$) and $\kappa_{c_{2,1}}$ ($\kappa_{c_{2,2}}$)
equal to the square of the first (second) coefficient in the right
side of Eq. (9) and Eq. (10), respectively. Set
$A_{\kappa_{1}}$ $=$ $A_{\kappa_{2}}$ $=$ $0$, decays from $|S\rangle$ to
$|11\rangle$ and from $|T\rangle$ to $|11\rangle$ can be both
largely suppressed. On the other hand, the microwave fields drive
the transition between the three states $|00\rangle$, $|T\rangle$
($|S\rangle$) and $|11\rangle$ for $\theta_{M}=0(\pi)$. The dynamics
of the full master equation in Fig. 3 (a) and
(b) illustrates that we can obtain state $|S\rangle$ or $|T\rangle$
of high fidelity, and the time needed for reaching the entangled
steady-state $|T\rangle$ is about two times as large as that of
$|S\rangle$. This is because that the optimal ratio of
$\kappa_{c_{1,1}}$/$\kappa_{c_{1,2}}$ is about $2$ times as large as
$\kappa_{c_{2,1}}$/$\kappa_{c_{2,2}}$.
\begin{figure}
\centering \subfigure[]{ \label{Fig.sub.a}
\includegraphics[width=0.8\columnwidth]{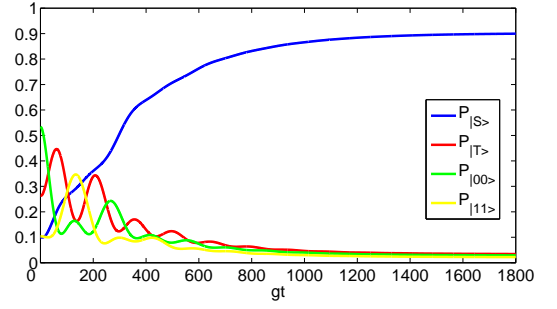}}
\subfigure[]{ \label{Fig.sub.b}
\includegraphics[width=0.78\columnwidth]{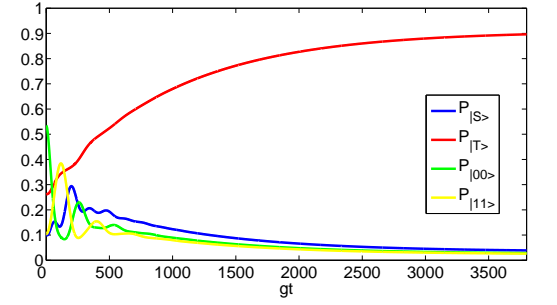}}
\subfigure[]{ \label{Fig.sub.c}
\includegraphics[width=0.81\columnwidth]{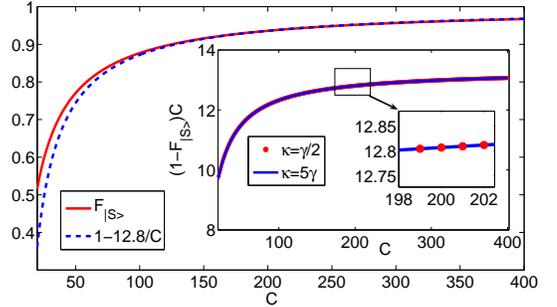}}
\caption{(Color online) The populations of four states $|S\rangle$,
$|T\rangle$, $|00\rangle$ and $|11\rangle$ versus the dimensionless
parameter $gt$ for a random initial state. Both curves are plotted
for $C=200$, $\kappa=\gamma/2$, $\Omega_{M}=2\Omega/5$,
$\Omega=g/20$ with $\Delta$, $\delta$ and $J$ being the optimal
values for two entangled steady-states. (a) $\theta_{M}=0$. (b)
$\theta_{M}=\pi$. (c) The fidelity $F_{|S\rangle}$ for steady-state
$|S\rangle$ versus $C$, and the coefficient of the linear scaling in
$F_{|S\rangle}$ as a function of $C$ with different ratios $\kappa/\gamma$ is
plotted in the inset.}
\end{figure}
The errors imposed by all possible atomic spontaneous emissions
should also be taken into account. We apply Eq. (6) again to derive
four analytic expressions of effective spontaneous emissions with
the other Lindblad operators $L^{\gamma_1}$, $L^{\gamma_2}$,
$L^{\gamma_3}$ and $L^{\gamma_4}$
\begin{eqnarray}\label{e12-e13}
L^{\gamma_1}_{eff}&=&\sqrt{\frac{\gamma}{2}}[\frac{\Omega}{2}|\widetilde{R}_{1}||00\rangle\langle00|
+\frac{\Omega}{4}|\widetilde{R}_{2}|(|T\rangle\langle
T|+|S\rangle\langle
T|)\cr&&+\frac{\Omega}{4}|\widetilde{R}_{3}|(|T\rangle\langle
S|+|S\rangle\langle S|)],\\
L^{\gamma_3}_{eff}
&=&\sqrt{\frac{\gamma}{2}}[\frac{\Omega}{2\sqrt{2}}|\widetilde{R}_{1}|(|T\rangle\langle00|+|S\rangle\langle00|)
\cr&&+\frac{\Omega}{2\sqrt{2}}(|\widetilde{R}_{2}||11\rangle\langle
T|+|\widetilde{R}_{3}||11\rangle\langle S|)],
\end{eqnarray}
where $|\cdot|$ denotes modulus of the symbol in it,
$L^{\gamma_2}_{eff}=L^{\gamma_1}_{eff}$ and $L^{\gamma_4}_{eff}=
L^{\gamma_3}_{eff}$. The operators of effective spontaneous emission
for $|S\rangle$ state are
\begin{eqnarray}\label{e14-e15}
L^{\gamma_1}_{eff,S}&=&L^{\gamma_2}_{eff,S}=\sqrt{\gamma_{S,i=1,2}}|T\rangle\langle
S|,\\
L^{\gamma_3}_{eff,S}&=&L^{\gamma_4}_{eff,S}=\sqrt{\gamma_{S,i=3,4}}|11\rangle\langle
S|,
\end{eqnarray}
and the operators of that for $|T\rangle$ state are
\begin{eqnarray}\label{e16-e17}
L^{\gamma_1}_{eff,T}&=&L^{\gamma_2}_{eff,T}=\sqrt{\gamma_{T,i=1,2}}|S\rangle\langle
T|,\\
L^{\gamma_3}_{eff,T}&=&L^{\gamma_4}_{eff,T}=\sqrt{\gamma_{T,i=3,4}}|11\rangle\langle
T|,
\end{eqnarray}
where
\begin{eqnarray}\label{e18}
\gamma_{eff} &\simeq&
\frac{(\frac{\gamma\Omega^2}{2})\{(gJ)^2+[\kappa(\Delta\delta-\frac{g^2}{2})+\gamma\frac{(\delta^2-J^2)}{2}]^2\}}
{(g^2-\Delta\delta)^2(g^4+\kappa^2\Delta^2)},\cr&&
\end{eqnarray}
and $\gamma_{S,i=1,2}$ $=$ $\gamma_{T,i=1,2}$ $=$ $\gamma_{eff}/16$,
$\gamma_{S,i=3,4}$ $=$ $\gamma_{T,i=3,4}$ $=$ $\gamma_{eff}/8$. Then
we use the rate equation to evaluate the fidelity for the state
$(j=S$ or $T)$
\begin{figure}[H]
\includegraphics[width=1\columnwidth]{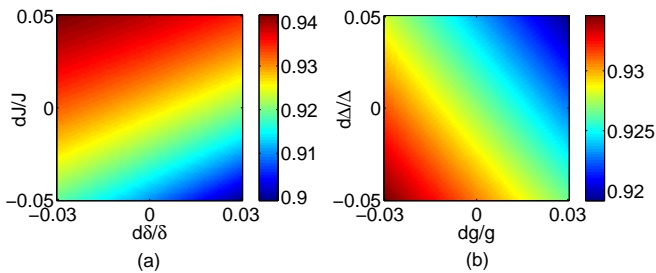}
\caption{(Color online) $F_{|S\rangle}$ in the effective two-qubit system versus fluctuations of
various parameters. (a) $F_{|S\rangle}$ vs $\frac{dJ}{J}$ and
$\frac{d\delta}{\delta}$; (b) $F_{|S\rangle}$ vs
$\frac{d\Delta}{\Delta}$ and $\frac{dg}{g}$.}
\end{figure}
\begin{eqnarray}\label{e19}
\dot{P}_{j}=\kappa_{a}P_{00}-(\kappa_{b}+\sum_{i=1}^{4}\gamma_{j,i})P_{j},
\end{eqnarray}
where $P_{j}$ is the probability to be in the state $j$. The first
term on the right side of Eq. (19) represents the population
decaying into the state $j$ with the rate $\kappa_a$, while the
other terms express the population leaking out of the state $j$ with
the rate $\kappa_{b}+\sum_{i=1}^{4}\gamma_{j,i}$. Suppose $P_{j}$
$\simeq$ $1$ and the probability in each of the other three states
is nearly $P_{00}$, then
\begin{eqnarray}\label{e20}
1-F_{|S\rangle}\approx(3\frac{g_{eff}^2\kappa}{C_{k_{1}}^{2}+D_{k_{1}}^{2}}+9\gamma_{eff})
/[\frac{(\delta+J)^{2}g_{eff}^2\kappa}{A_{k_{1}}^{2}+B_{k_{1}}^{2}}],\cr
\end{eqnarray}
where $F_{|S\rangle}$ $=$ $|\langle S|\rho_{SS}|S\rangle|$ is the
fidelity of state $|S\rangle$. Setting $\delta g^2=\Delta(\delta^2-J^2)$ and
$\kappa(\Delta\delta-g^2/2)$ $\simeq$ $\gamma(\delta^2-J^2)/2$, the optimal fidelity of the
entanglement can be obtained. The effective two-qubit system in the
inset of Fig. 3 (c) shows that the fidelity scaling of state
$|S\rangle$ is independent of different ratios $\kappa/\gamma$, then
we find out the actual constants for maximizing the fidelity as
follows
\begin{eqnarray}\label{e21}
1-F_{|S\rangle}\approx12.8 C^{-1}.
\end{eqnarray}

The influences of different parameter fluctuations on the fidelity
$F_{|S\rangle}$ of entangled state are considered. As shown in Fig.
4 (a) and (b), $F_{|S\rangle}$ keeps above $90\%$ even $5\%$
fluctuations in these parameters. The preparation process of state
$|T\rangle$ is similar to that of $|S\rangle$.

Photonic band gap cavities coupled to atoms or quantum dots are
suitable candidates for realizing the proposal. Cooperativity of
value $C$ $\sim$ $100$ has been realized \cite{Nature2007-445-896}.
The cavity modes can be coupled via the overlap of their evanescent
fields or via an optical fiber, and photon hopping between two
cavities has been observed \cite{PRB2000-61-R11855}.

In conclusion, we have proposed a scheme for dissipative preparation
of entanglement between two atoms that are distributed in two
coupled cavities. We find the linear scaling of the fidelity is a
quadratic improvement compared with distributed entangled state
preparation protocols based on unitary dynamics.

L.T.S., X.Y.C., H.Z.W, and S.B.Z acknowledge support from the
National Fundamental Research Program Under Grant No. 2012CB921601,
National Natural Science Foundation of China under Grant No.
10974028, the Doctoral Foundation of the Ministry of Education of
China under Grant No. 20093514110009, and the Natural Science
Foundation of Fujian Province under Grant No. 2009J06002. Z.B.Y is
supported by the National Basic Research Program of China under
Grants No. 2011CB921200 and No. 2011CBA00200, and the China
Postdoctoral Science Foundation under Grant No. 20110490828.


\end{document}